\documentclass[aps,prl,showpacs,twocolumn,amsmath,amssymb]{revtex4-1}

\usepackage{graphicx}
\usepackage{dcolumn}
\usepackage{bm}

\preprint{APS/123-QED}

\begin{document}

\title{Observation of local temporal correlations in trapped quantum gases}
\author{Vera Guarrera}
\email{guarrera@physik.uni-kl.de}
\author{Peter W\"{u}rtz}
\author{Arne Ewerbeck}
\author{Andreas Vogler}
\author{Giovanni Barontini}
\author{Herwig Ott}
\affiliation{Research Center OPTIMAS, Technische Universit\"at Kaiserslautern, 67663 Kaiserslautern, Germany}

\date{\today}

\begin{abstract}
We measure the temporal pair correlation function $g^{(2)}(\tau)$ of a trapped gas of bosons above and below the critical temperature for Bose-Einstein condensation. The measurement is performed {\it in situ} using a local, time-resolved single-atom sensitive probing technique. Third and fourth order correlation functions are also extracted. We develop a theoretical model and compare it with our experimental data, finding good quantitative agreement and highlighting the role of interactions. Our results promote temporal correlations as new observables to study the dynamics of ultracold quantum gases. 
\end{abstract}


\maketitle
The intriguing effect of particle bunching was first observed in a seminal experiment by Hanbury Brown and Twiss (HBT), where they studied correlations among pairs of photons coming from a chaotic source \cite{hbt}. The result, which could be explained in terms of classical waves, had a difficult route to be accepted under a particle perspective. The full quantum theory, due to Glauber \cite{glauberphotons}, signed the birth of quantum optics and made the formalism, with all its physical content, available for massive particles. Over the years, analogous HBT experiments were performed with electrons \cite{electrons}, neutrons \cite{neutrons} and cold atoms \cite{shimizu,aspect}. The possibility to extract information about the quantum statistics and the coherence, paved by the HBT experience, conjugated with the capabilities of deriving the temperature and the spatial order \cite{bloch}, contributed to make this technique one of the most powerful to probe atomic systems. The quest to understand the behavior of more and more complex samples suggests its application to the study of strongly interacting 1D gases \cite{shlyapnikov, weiss}, disordered \cite{disorder} and supersolid phases \cite{supersolid} and to identify non trivial excitations \cite{balbinot}.
While first order correlations are often accessible via interference experiments, higher order correlations require in general the recording of density or atom number fluctuations by a probe sensitive enough to detect single particles (counting techniques) or, at least, atomic shot-noise (absorption imaging). In order to have a good statistical description, an average over many realizations of the system (in theory all possible realizations) is needed.
Consequently correlations, especially at orders higher than two, are usually difficult to measure because of the huge statistics required for a reliable signal. Only in some limited cases, intrinsic processes in a quantum gas such as photoassociation or three body losses can be used as a sensitive probe for higher order correlations at zero distance \cite{weiss2,cornell}. The direct observation of third order correlations is still challenging and, using standard techniques, requires a considerable effort in data collection and analysis \cite{australianscience}.
\begin{figure}[t!]
\begin{center}
\includegraphics[width=0.45\textwidth]{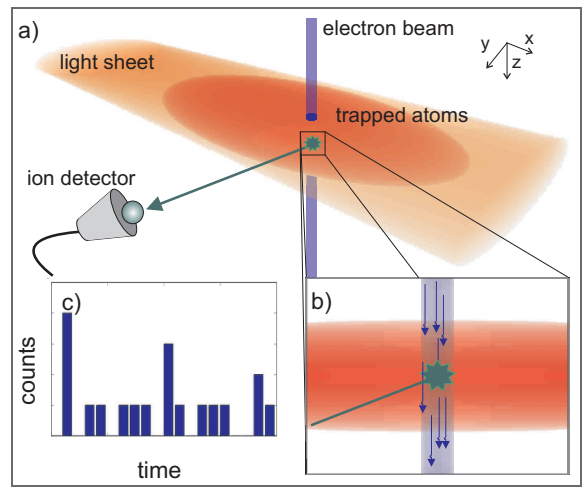}
\end{center}
\caption{(Color online) a) Schematic of the apparatus. The electrons of the focussed beam collide with atoms leading to ionization with a probability of $40\%$. The ions are guided to the detector to be counted. The average count rate is $1.7$ kHz. b) The electron-atom interaction volume is defined by the electron beam itself and by the extension of the cloud along the $z$ axis. c) A typical measurement of the ion signal showing the first $20$ bins.} \label{fig:setup}
\end{figure} 
In this, like in the great majority of the above mentioned experiments with ultracold gases, correlations have only been studied in the spatial domain whereas the temporal counterpart has been very poorly explored, limited only to the characterization of atomic beams \cite{shimizu,esslinger}.
Boosted by recent achievements \cite{Gericke2008, greiner, kuhr}, spatially resolved single atom sensitive detection methods can give direct access to higher order correlation functions {\it in situ}. Such experiments are not complicated by the time of flight expansion and directly manifest the properties of the many-body system. Moreover if the detection method is only locally probing the system, even time-resolved correlation measurements are possible. 

In this Letter, we report on the first observation of {\it temporal} thermal bunching of ultracold bosonic atoms in a trap. Using scanning electron microscopy as time-resolved local detection method, our measurements directly yield the second, third and fourth order time correlation functions. Notably, due to the spatially resolved measurement and due to the possibility to follow the dynamics of the system, our technique is effectively an original method for characterizing ultracold quantum systems, both in space and time. All these features can turn out to be extremely useful in providing deep insight into the dynamics of strongly correlated many-body quantum systems \cite{1d}.

The general form of the normalized spatio-temporal correlation function of $n$ particles at position $\textbf{r}_i$ at time $t_i$, with $i=1,...,n$, is given by: 
\begin{eqnarray}
&&g^{(n)}(\textbf{r}_1,t_1;...;\textbf{r}_n,t_n)= \nonumber \\
&&\frac{\langle \widehat{\Psi}^\dagger(\textbf{r}_1,t_1)... \widehat{\Psi}^\dagger(\textbf{r}_n,t_n) \widehat{\Psi}(\textbf{r}_n,t_n)...  \widehat{\Psi}(\textbf{r}_1,t_1) \rangle}{\langle \widehat{\Psi}^\dagger(\textbf{r}_1,t_1) \widehat{\Psi}(\textbf{r}_1,t_1)\rangle...\langle \widehat{\Psi}^\dagger(\textbf{r}_n,t_n)\widehat{\Psi}(\textbf{r}_n,t_n)\rangle}
\label{eq:g2 general}
\end{eqnarray}
where $\widehat{\Psi}$ are the bosonic operators and $\langle ... \rangle$ indicates the ensemble average.
We first derive an analytical expression of $g^{(1)}(\textbf{r}_1,t_1;\textbf{r}_2,t_2)$ for an ideal Bose gas at temperature $T$, trapped in a harmonic potential $V(\textbf{r})=m\omega^2r^2/2$, with average trapping frequency $\omega$, extending the approach of Ref. \cite{naraschewski} to take into account also the temporal evolution.
Given $\tau=t_2-t_1$ and $\textbf{r}=\textbf{r}_2-\textbf{r}_1$ and assuming $\omega\tau,  \hbar\omega/(k_BT)\ll1$, we obtain:
\begin{equation}
g^{(1)}(\textbf{r},\tau)=\frac{1}{\left(1+i \frac{\tau}{\tau_c} \right)^{3/2}}\exp \left( -\frac{m r^2}{2\hbar \tau_c^2}\frac{\tau_c+i\tau}{1+\left(\frac{\tau}{\tau_c}\right)^2}\right)
\label{eq:g1}
\end{equation}
where $\tau_c=\frac{\hbar}{k_B T}$ is defined as the correlation time. 
From the above expression we can derive any higher order correlation function for thermal bosons and, in particular, the second order correlation can be easily calculated as $g^{(2)}(\textbf{r},\tau)=1+\vert g^{(1)}(\textbf{r},\tau) \vert^2 $.
The time correlation function $g^{(2)}(0,\tau)$ can be interpreted as the probability to detect a particle a time $\tau$ after another particle at the same position ($r=0$). For a thermal cloud of bosons this function decreases from $2$ to $1$ on a time-scale related to the correlation time $\tau_c$. A value of the pair correlation function higher than $1$ indicates bunching of thermal bosons in time. For a coherent source like a BEC, we expect instead $g^{(2)}(0,\tau)=1$ for any $\tau$ \cite{naraschewski}, i.e. a flat detection probability in $\tau$.

Since our setup has been described earlier \cite{gericke2007,wurtz}, we briefly illustrate the experimental procedure we have followed to measure the atomic correlation functions.
The key feature of our experiment is a scanning electron microscope which is implemented on a standard apparatus for the production of ultracold quantum gases. The measurement principle is based on the electron impact ionization of the atoms with subsequent ion detection. A sketch of the working principle is depicted in Fig.\ref{fig:setup}. In addition to a spatial resolution of better than $150$ nm, the technique is characterized by a sequential detection method. Thus, time-dependent quantities such as the second or higher order correlation function $g^{(n)}(\tau)$, $n=2,3,..$, become experimentally accessible.
\begin{figure}[t!]
\begin{center}
\includegraphics[width=0.475\textwidth]{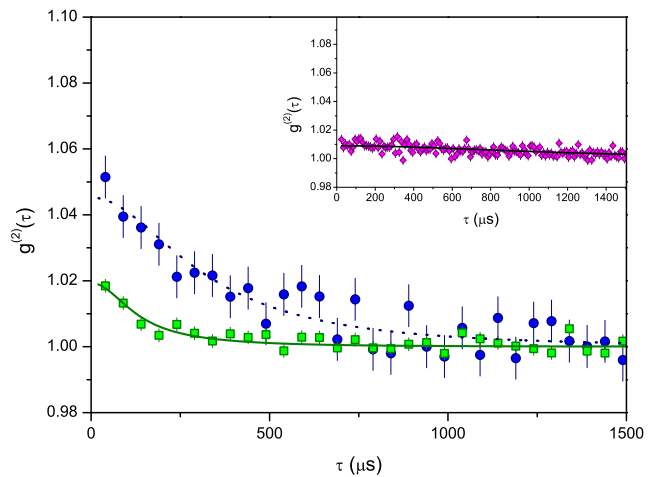}
\end{center}
\caption{(Color online) Normalized second order temporal correlation function. Data (dots) acquired at different temperatures (circles for $45$ nK, squares for $100$ nK) are plotted together with the fitting functions (lines) explained in the text. The inset displays the data (diamonds) acquired for a BEC and the corresponding fit (line). Please note that even well below the critical temperature (the thermal fraction cannot be detected in time of flight absorption imaging) we are able to measure a small residual bunching induced by the thermal component.
} \label{fig:g2}
\end{figure}
To prepare the atomic sample we load approximately $1.5\times 10^6$ $^{87}$Rb atoms from a magneto-optical trap in a CO$_2$ dipole trap. The atoms are then evaporatively cooled above or below the critical temperature. We can typically prepare cold thermal samples of $2-3 \times 10^5$ atoms at $150$ nK as well as BECs of up to $1 \times 10^{5}$ atoms in the $F=1$ hyperfine ground state. Since integration along the probing line reduces the correlation signal amplitude (see for example \cite{aspect}), after the evaporation we compress the cloud along the direction of the electron beam by adiabatically transfering it to a \emph{light sheet} dipole trap. This trap is realized by means of a focussed elliptical $852$ nm laser beam, with waists $(6.5, 130)$ $\mu$m. We then completely switch off the CO$_2$ trap and the sample is held by the light sheet alone for all the duration of the measurements (Fig. \ref{fig:setup}). The final frequencies, for a beam power of $5$ mW, are $\omega_{x,y,z} \simeq 2\pi \times (13, 27, 580)$ Hz.  
To detect the atoms, an electron beam of $6$ keV energy, $20$ nA current and $120$ nm FWHM is focussed at the centre of the cloud, where the atom density and hence the number of ions produced is maximal. This choice is only meant to allow for a reduced number of experimental cycles but any other point of the cloud can be investigated as well. The electron-atom interaction is confined within a distance smaller ($x,y$ directions) or comparable ($z$ direction) to the typical correlation lengths of our atomic samples. The ions produced by electron impact ionization are collected in a channeltron. For each bin of $10$ $\mu$s we record the number of ions detected and the absolute detection time, as shown in the inset of Fig.\ref{fig:setup}. In a typical measurement of $600$ ms duration we extract $\sim1000$ ions. 
After the measurement, the remaining atoms are released from the trap and imaged after time-of-flight (TOF) to extract the temperature. 
In order to optimize signal-to-noise ratio we compute the correlation functions over about $1500$ repetitions of the experiment.

From Eq.(\ref{eq:g2 general}), integrating on the absolute time and averaging on the different repetitions of the experiment, we calculate the correlation function $g^{(2)}(\tau)$ for a cold thermal cloud. In Fig.\ref{fig:g2} we report the measurements for two different temperatures $T=45$ nK and $T=100$ nK. In both cases, we can clearly observe bunching. In these data plots we have omitted the first two bins, which show an abnormally high value of $g^{(2)}(\tau)$. These points are affected by extra ion counts introduced by the electronics of the detector. As a test for the procedure, we apply the same technique to a reference measurement which is \emph{a priori} uncorrelated. This is obtained by probing a thermal cloud at $T=230$ nK with an electron beam of waist well larger than the correlation length. As expected, no bunching signal is detectable in the measurement, making sure that what we have observed in the data sets taken in standard conditions is genuine thermal bunching.
 
Averaging over several experimental cycles, fluctuations in the total atom number affect the normalization of the correlation function. As a result, an offset shifts the uncorrelated signal to a value $1\%$ above $1$. To compensate for these fluctuations we normalize $g^{(2)}(\tau)$ by the factor $1+(\sigma^2-\langle N \rangle )/ \langle N \rangle ^2$, where $\sigma^2$ and $\langle N \rangle$ are respectively the variance and the mean value of the total atom number in the different experimental realizations.
\begin{figure}[t!]
\begin{center}
\includegraphics[width=0.45\textwidth]{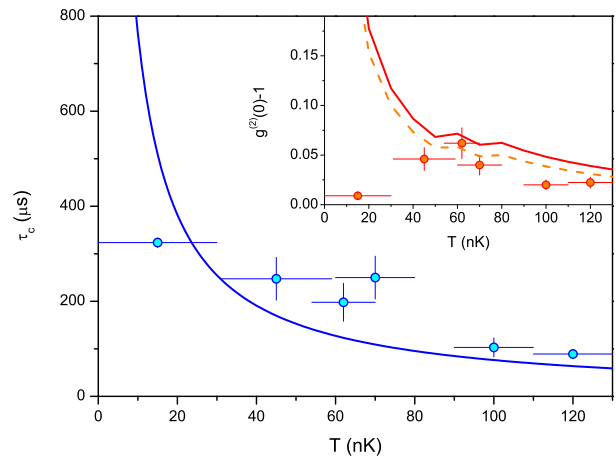}
\end{center}
\caption{(Color online) Correlation time $\tau_c$ as a function of the temperature. The values obtained fitting the experimental data (circles) are compared with the function $\tau_c=\hbar/(k_B T)$ (line). The inset shows the fitted bunching enhancements (circles) and the corresponding theoretically expected values for the non-interacting (solid line) and the interacting (dashed line) model.} \label{fig:tauc}
\end{figure}
A measurement with a BEC is also presented in the inset of Fig.\ref{fig:g2}. Notably, we can still detect a small correlation signal due to a residual thermal fraction present well below the critical temperature $T_c$.
The $g^{(2)}(\tau)$ data points are fitted with the function we derived for an ideal non interacting gas of bosons integrated over the volume, leaving the amplitude and $\tau_c$ as free parameters (lines in Fig.\ref{fig:g2}). The results of the fits are shown in Fig.\ref{fig:tauc}, where the extracted correlation times are plotted as a function of the temperature of the sample. Since we cannot derive the temperature for a cloud with more than $70\%$ condensate fraction by TOF measurements, in the plots of Fig.\ref{fig:tauc} we indicate as the BEC temperature the average temperature between zero and the coolest measurable temperature of $30$ nK. The agreement between the experimental data and the theoretical function $\tau_c=\hbar/(k_BT)$ is fairly good. Hence this technique can be proposed as a local probe for the temperature of ultracold samples, especially when standard imaging techniques fail.

\begin{figure}[t!]
\begin{center}
\includegraphics[width=0.45\textwidth]{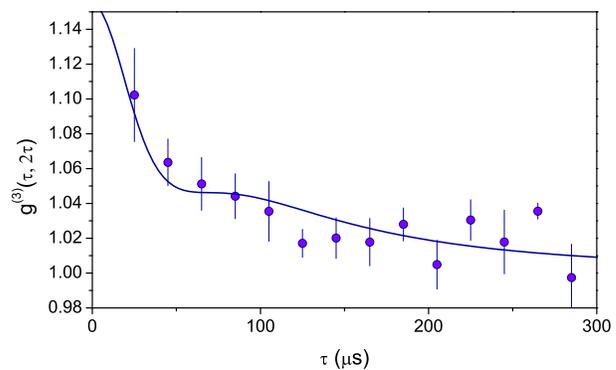}
\end{center}
\caption{(Color online) Normalized third order correlation function (dots) along the axis ($\tau_1=t_2-t_1,\tau_2=t_3-t_1=2\tau_1$) for a thermal cloud with $T=100$ nK. The solid line is the fitting function explained in the text.} \label{fig:g3}
\end{figure}


In the inset of Fig.\ref{fig:tauc} we show the fitted amplitude of the normalized second order  correlation functions at different temperatures together with the values expected from the volume integration of the non-interacting model and of its extension to the interacting case \cite{naraschewski}:
\begin{equation}
g_{int}^{(2)}(\textbf{r},0)=1+\frac{2a^2}{r^2}+\vert g^{(1)}(\textbf{r},0)\vert^2 \left(1-\frac{4a}{r}, \right)
\label{eq:g2int}
\end{equation}
being $a$ the s-wave scattering length. 
As expected \cite{naraschewski}, repulsive interactions play a significant role for \emph{in situ} measurements since they induce short-range anti-bunching that reduces or overcomes the bunching signal of bosons. Our measurements are well reproduced by the interacting model. In the regime where the BEC fraction is dominant, the signal is further reduced by the presence of a large coherent component in the integrating volume and by the role of interactions in the condensate. 

In addition, we also calculate the third order correlation function, extracting it from the same data sets for which we derived the $g^{(2)}(\tau)$. Bunching is expected to be more pronounced at higher orders $n$ as a consequence of the factorial law $n!$ that regulates the dependence of the correlated to the uncorrelated amplitudes. For this reason higher order correlations can be employed as a highly sensitive test for coherence, with the only drawback represented by the need of high statistics, as proved by the very few experiments reporting correlations at orders higher than $2$.

In Fig.\ref{fig:g3} we show an example of the third order correlation function $g^{(3)}(\tau_1,\tau_2)$, obtained for the data set at $T=100$ nK, along the axis $(\tau_1, \tau_2=2\tau_1)$. The solid line in the picture is a fit made with the volume integration of the non interacting model
$g^{(3)}(\tau_1,\tau_2)=1+\vert g^{(1)}(\tau_1)\vert^2 + \vert g^{(1)}(\tau_2)\vert^2 +\vert g^{(1)}(\tau_2 -\tau_1)\vert^2
+2\Re(g^{(1)}(\tau_1)g^{(1)}(-\tau_2)g^{(1)}(\tau_2-\tau_1))$,
along the same axis, leaving the amplitude as the only free parameter.
The measured amplitudes of the third order correlation functions $g^{3}(0,0)-1$ are $0.14\pm0.07$ and $0.10\pm0.02$, respectively for the data sets at $T=45$ nK and $T=100$ nK. Our measurements show that the amplitudes of the $g^{(3)}$ correlations are also affected by repulsive interactions since the bare volume integration of the non-interacting model is not sufficient to explain the observed reduction of the signal. Interestingly, the corresponding ratios $(g^{3}(0,0)-1)/(g^{2}(0)-1)$, which are $3.2\pm2.4$ and $5.0\pm2.2$, show instead agreement with the values derived from the non-interacting model: $4.2$ and $2.8$. This result is in accordance with the seminal work on the three-body losses \cite{cornell} and may signal that the effect of interactions does not scale with the order of the correlation function. 
Finally we measured the fourth order correlation function amplitude: for $T=45$ nK, $g^{(4)}(0,0,0)-1= 0.46 \pm 0.42$ and for $T=100$ nK, $g^{(4)}(0,0,0)-1= 0.23 \pm 0.05$. At times longer than $\tau_c$, the $g^{(4)}$ correlation signals drop respectively to the values of $0.8\pm0.1$ and $1.03\pm0.01$. The evaluation of $g^{(4)}$ correlations turns out to be strongly affected by the poor statistics, thus rendering any comparison with theory difficult. Further investigation on fourth and higher order correlations can be, however, useful to identify the influence of interactions in multiple-particle bunching.

In summary, we have observed temporal pair correlations in a cold gas above $T_c$. We have measured $g^{(2)}(\tau)$ for different temperatures of the atoms and compared the results with a theoretical model, that we derived for a non-interacting and for an interacting system of thermal bosons. A measurement on a BEC has also revealed a minimal bunching compatible with the presence of a residual thermal fraction. A signal of third and fourth order correlations has been derived from the same data. The amplitudes of the bunching enhancement of $g^{(3)}$ and $g^{(2)}$ correlations are affected by repulsive interparticle interactions, while their ratios are compatible with a non-interacting modellization. The electron microscopy technique, which we use, represents a powerful, fast and efficient way to measure correlations, with the possibility to probe the system in space and time. Notably, pair correlations in time can give access to the dynamical structure factor and reveal, in contrast to static pair correlations, the dynamical properties of the quantum system.  

\begin{acknowledgments}
We acknowledge financial support by the DFG within the SFB/TRR 49 and GRK 792. V. G. and G. B. are supported by a Marie Curie Intra-European Fellowship.
\end{acknowledgments}

\end{document}